\documentclass[a4paper,12pt]{article}

\usepackage[english]{babel}
\usepackage[a4paper,tmargin=3truecm,bmargin=3truecm,rmargin=2.5truecm,
lmargin=2.5truecm,twoside,verbose=true]{geometry}

\usepackage{cancel,graphicx}
\usepackage[dvips]{hyperref}

\usepackage{amsmath,amssymb}
\usepackage{slashed}
\numberwithin{equation}{section}

\allowdisplaybreaks[1]

\renewenvironment{thebibliography}[1]
         {\section*{References}\frenchspacing\small
          \begin{list}{[\arabic{enumi}]}
         {\usecounter{enumi}\parsep=2pt\topsep 0pt
         \settowidth{\labelwidth}{[#1]}
         \leftmargin=\labelwidth\advance\leftmargin\labelsep
         \rightmargin=0pt\itemsep=1pt\sloppy}}{\end{list}}

\title{Noncommutative Induced Gauge Theories \\ on Moyal
Spaces\footnote{Talk given at the "International Conference on Noncommutative Geometry and Physics", April  2007, Orsay (France). Work
supported by ANR grant NT05-3-43374 ``GenoPhy''.}}
\author{Jean-Christophe Wallet}
\date{}

\begin{document}

\maketitle
\vspace*{-1cm}
\begin{center}
\textit{Laboratoire de Physique Th\'eorique, B\^at.\ 210\\
    Universit\'e Paris XI,  F-91405 Orsay Cedex, France\\
    e-mail:
\texttt{jean-christophe.wallet@th.u-psud.fr}}\\[1ex]

\end{center}%

\vskip 2cm

\begin{abstract}
Noncommutative field theories on Moyal spaces can be conveniently handled within a framework of noncommutative geometry. Several renormalisable matter field theories that are now identified are briefly reviewed. The construction of renormalisable gauge theories on these noncommutative Moyal spaces, which remains so far a challenging problem, is then closely examined. The computation in 4-D of the one-loop effective gauge theory generated from the integration over a scalar field appearing in a renormalisable theory minimally coupled to an external gauge potential is presented. 
The gauge invariant effective action is found to involve, beyond the expected noncommutative version of the pure Yang-Mills action, additional terms that may be interpreted as the gauge theory counterpart of the harmonic term, which for the noncommutative $\varphi^4$-theory on Moyal space ensures renormalisability. A class of possible candidates for renormalisable gauge theory actions defined on Moyal space is presented and discussed.
\end{abstract}

\pagebreak

\section{Introduction.}

An intense activity has been devoted to the
study of various field theories defined on Moyal spaces
(see e.g.\ \cite{Douglas:2001ba,Wulkenhaar:2006si}). These noncommutative field theories involve most of the features stemming from noncommutative geometry \cite{CONNES,CM} and are thus interesting
in themselves. This interest was further increased by the observation that somewhat similar noncommutative field theories might emerge from some limiting regime of string theory and matrix theory in
magnetic backgrounds \cite{Seiberg:1999vs} (See also
\cite{Witten:1985cc}). In this talk, I will report recent progress \cite{PAPER1} toward the construction of renormalisable actions for gauge theories defined on noncommutative Moyal spaces which so far remains an unsolved and challenging problem. The corresponding results have been obtained in collaboration with A. de Goursac and R. Wulkenhaar \cite{PAPER1}. Our results and conclusions agree with those obtained recently by H. Grosse and M. Wohlgennant, as reported in \cite{Grosse:2007hh}. In noncommutative geometry, the commutative algebras of functions defined on differentiable manifolds (which may be viewed as modelling the coordinates spaces)
are replaced by associative algebras interpreted as algebras of functions on ``noncommutative spaces''.
Within this algebraic framework, natural noncommutative analogues of
the main geometrical objects usually involved in field theories can be defined thus opening the way for the construction of noncommutative analogues
of field theories (see e.g.\ \cite{Gayral:2006wu}). The relevant configuration spaces for the noncommutative
field theories are modules over the associative algebras, naturally interpreted as noncommutative analogues for the set of sections of vector bundles, while the linear spaces of the derivations of these algebras can be viewed as the noncommutative counterparts of the spaces of vector fields of commutative spaces . This finally permits one to define natural noncommutative extensions of connections and curvatures \cite{CONNES,CM}. One example of associative algebra, among many others, is provided by the associative Moyal algebras \cite{Gracia-Bondia:1987kw,Varilly:1988jk} which will then play the role of ``noncommutative Moyal spaces''. Note that Moyal algebras have been considered in the mathematical literature a long time ago from various viewpoints (see e.g. \cite{Grossmann:1968} and references therein).\par
It is known that the simplest generalisations of scalar theories to Moyal space suffer from the UV/IR-mixing
\cite{Minwalla:1999px,Chepelev:1999tt}, which makes the
renormalisability very unlikely. Recall that the UV/IR-mixing results
from the existence of potentially dangerous non-planar diagrams which,
albeit UV finite, become singular at exceptional low external
momenta. This generates UV divergences in higher
order diagrams in which they are involved as subdiagrams, signalling that 
that UV and IR scales are non-trivially related which should invalidate a Wilson-type 
renormalisation scheme \cite{Wilson:1973jj,Polchinski:1983gv}. An appealing solution to the
UV/IR-mixing has been recently proposed by Grosse and Wulkenhaar
\cite{Grosse:2004yu,Grosse:2003aj} within the noncommutative
$\varphi^4$ model ($\varphi$ real-valued) on the 4-dimensional Moyal 
space. They showed that the UV/IR-mixing can be suppressed by
supplementing the initial action with a harmonic oscillator term
leading to a renormalisable quantum field theory. The
initial beautiful proof \cite{Grosse:2004yu} performed within the
so-called matrix-base formalism was further simplified through a reformulation into the (position) $x$-space
formalism in \cite{Gurau:2005gd}. Other renormalisable noncommutative matter field theories on Moyal
spaces have then been identified. One is the complex-valued scalar theory
studied in \cite{Gurau:2005gd} which is nothing but a modified
version of the LSZ model \cite{Langmann:2003if,Langmann:2003cg} {\footnote{Note that the
scalar theory in \cite{Grosse:2003nw} is super-renormalisable. Besides, 
interesting solvable noncommutative scalar field theories have
also been considered in
\cite{Grosse:2005ig,Grosse:2006qv,Grosse:2006tc}}}. As far as fermionic
theories are concerned, a Moyal space version of the Gross-Neveu model
\cite{Gross:1974jv}),
called the orientable noncommutative Gross-Neveu model, has been
recently shown to be renormalisable to all orders
\cite{Vignes-Tourneret:2006nb} (see also
\cite{Lakhoua:2007ra}). This latter noncommutative theories still exhibit some residual UV/IR-mixing, even in the presence of the suitable extension of the harmonic term of \cite{Grosse:2004yu}, which however does not prevent the theory from being {\it{perturbatively}} renormalisable. This indicates that further investigations are needed to clarify the actual role of various generalisations of the above-mentioned harmonic oscillator term, of the related covariance under the Langmann-Szabo duality \cite{Langmann:2002cc} and of their impact in the control of the UV/IR-mixing and renormalisability.\par

So far, the construction of a renormalisable gauge theory on
Moyal spaces remains still unsolved and appears to be a quite challenging problem. The naive noncommutative extension of the pure Yang-Mills action on Moyal
spaces has UV/IR mixing \cite{Hayakawa:1999yt,Matusis:2000jf} which makes its 
renormalisability quite unlikely unless it is suitably modified. Unfortunately, the "harmonic solution" proposed
in \cite{Grosse:2004yu} cannot be merely extended to gauge theories on Moyal spaces. Finding such a suitable extension would first amount to determine whether or not the naive noncommutative Yang-Mills action can be supplemented by additional gauge invariant terms providing a natural "gauge theory counterpart" of those harmonic terms. As shown in \cite{PAPER1} and as it will be explained during this talk, this can be conveniently achieved by computing at the one-loop order the noncommutative effective gauge theory action obtained by integrating out the matter degree of freedom of a matter field theory with harmonic term minimally coupled to an external gauge potential. The calculation performed within the $x$-space formalism singles out \cite{PAPER1} a class of gauge invariant actions given by
\begin{align}
S=\int d^4x \Big(\frac{\alpha}{4g^2}F_{\mu\nu}\star F_{\mu\nu}
+\frac{\Omega^\prime}{4g^2}\{{\cal{A}}_\mu,{\cal{A}}_\nu\}^2_\star
+\frac{\kappa}{2} {\cal{A}}_\mu\star{\cal{A}}_\mu\Big) \label{eq:decadix1}
\end{align}
that may well involve (or may be used as a starting point to construct) suitable candidates for renormalisable actions for gauge theory defined on Moyal spaces. A similar investigation has been carried out independently by H. Grosse and M. Wohlgenannt \cite{Grosse:2007hh}. The corresponding basic ingredients (gauge transformations and starting matter action) and the computational tools (related to the matrix basis) are different from ours. But the resulting effective action and the related conclusions agree completely with ours. In \eqref{eq:decadix1}, ${\cal{A}}_\mu$ denotes a specific gauge covariant tensorial form, linked with the existence of a canonical gauge invariant connection \cite{Dubois-Violette:1989vq, Masson:1999} appearing naturally within the present noncommutative framework. It is related to the so called covariant coordinates in the string theory literature (see e.g \cite{Douglas:2001ba}). The 2nd term term built from the anticommutator and the 3rd term may be viewed as "gauge counterparts'' of the
harmonic term introduced in \cite{Grosse:2004yu} ($\alpha$, $\Omega^\prime$, $\kappa$ and $g$ are real parameters). Furthermore, the 3rd term involves a mass term for the gauge potential while such a bare mass term for a gauge potential is forbidden by gauge invariance in commutative Yang-Mills theories. It turns out that the presence of these quadratic and quartic terms in ${\cal{A}}_\mu$ is reflected in a non-vanishing vacuum expectation value for the gauge potential $A_\mu$, a rather unusual feature for a Yang-Mills type theory. Note that non trivial vacuum configurations have been studied recently in \cite{GOURSAC-TANASA-WALLET} within scalar models with harmonic term. The consequences of this non-trivial vacuum for the above class of gauge theories remain to be understood and properly controlled in view of a
further gauge-fixing of a (classical) gauge action stemming from \eqref{eq:decadix1} combined with a suitable regularisation scheme. 

\section{The Moyal algebra.}
In this section, I collect some basic mathematical tools involved in
the definition of the Moyal algebra. The Moyal product and the Moyal spaces are deeply rooted in Quantum
Mechanics through the operator calculus introduced by H. Weyl
\cite{WEYL}. The subsequent formulation of Quantum Mechanics on a
classical phase space was carried out in \cite{GROEN,MOYAL} thanks
to the introduction of a twisted product, hereafter called the Moyal product 
product, together with the introduction of some analog of the
quantum mechanical commutation relations, called the Moyal brackets. For more mathematical details, see e.g 
\cite{Grossmann:1968}, \cite{Gracia-Bondia:1987kw,Varilly:1988jk} and the references therein. Note that the Moyal spaces can be related to spectral triples, as shown in \cite{GGIS}. In the following, I will deal with four-dimensional Moyal spaces.
The extension to arbitrary (even) dimensional spaces is trivial. Let ${\cal{S}}$ and ${\cal{S}}^\prime$ denote respectively the space of complex-valued Schwartz functions on
${\mathbb{R}}^4$ with fast decay at infinity and the space of tempered distributions on ${\mathbb{R}}^4$. Let
$\Theta_{\mu\nu}$ be the invertible constant skew-symmetric matrix
\begin{align}
{{\Theta}}=\theta\begin{pmatrix} J &0 \\ 0& J \end{pmatrix} , \ J =\begin{pmatrix} 0&-1 \\ 1& 0 \end{pmatrix}\label{eq:theta}
\end{align}
where $\theta$ has mass dimension $-2$. The associative Moyal product, denoted hereafter by a $\star$ symbol,
can be first defined on ${\cal{S}} \times {\cal{S}}$ by{\footnote{
$\Theta.k\equiv\Theta_{\mu\nu}k^\nu$ and ${\hat{X}}$ is the Fourier transform of $X$}}
\begin{align}
(f\star h)(x)=&{{1}\over{(2\pi)^4}}\int d^4y\,d^4k\
f(x+{{1}\over{2}}\Theta.k)\,h(x+y)e^{i k.y} \nonumber\\
=&\frac{1}{\pi^4\theta^4}\int d^4yd^4z f(x+y)h(x+z)e^{-i2y\Theta^{-1}z}\nonumber\\
=&\frac{1}{(2\pi)^8}\int d^4k_1d^4k_2{\hat{f}}(k_1){\hat{h}}(k_2-k_1)e^{ik_2x}e^{-\frac{i}{2}k_1\Theta k_2},\label{eq:moyal}
\end{align}
for any $f,h\in{\cal{S}}$ such that $(f\star h)$$\in$$ {\cal{S}}$. From \eqref{eq:moyal}, one easily verifies that $(f\star g)\star h$$=$$f\star(g\star h)$, $f\star g$$\ne$$ g\star f $ and 
$(f\star h)^\dag$$=$$h^\dag\star f^\dag$ where $^\dag$ denotes the complex conjugation, $\forall f,g,h$$\in$${\cal{S}}$. One has the tracial property
\begin{align}
\int d^4x (f\star g)(x)=\int d^4x f(x).g(x) \label{eq:tracial}
\end{align}
where the symbol ``.'' is the (commutative) usual
product among functions. This, combined with the associativity of the $\star$-product  implies the cyclicity property
\begin{align}
\int d^4x (f_1\star f_2\star f_3)(x)=\int d^4x (f_2\star f_3 \star f_1)(x)=\int d^4x (f_3\star f_1\star f_2)(x)\quad \forall f_1,f_2,f_3\in{\cal{S}}
\end{align}
${\cal{S}}$ is not large enough to be physically interesting. In particular, it has
no unit. However, it can be enlarged to an algebra of tempered distributions using duality
of linear spaces such that the associativity of the $\star$-product and
the involution inherited from the complex
conjugation are preserved (see e.g \cite{Gracia-Bondia:1987kw,Varilly:1988jk}). The
$\star$-product is first extended to ${\cal{S}}^\prime \times{\cal{S}}$
and to ${\cal{S}} \times {\cal{S}}^\prime$ using the duality of linear spaces: one defines $T\star f$ and $f\star T$ through
$\langle T\star f,h \rangle$$ = $$\langle T,f\star h\rangle$, $\langle f\star T,h \rangle $$= $ 
$ \langle T,h\star f\rangle$, 
 $\forall f,h$$\in$${\cal{S}}$, $\forall T$$\in$$ {\cal{S}}^\prime$
where the duality bracket is $\langle f,g\rangle$$\equiv$$\int d^4x
f(x).g(x)$. Thanks to the smoothening
properties of the $\star$-product together with \eqref{eq:tracial}, $(T\star f)$$\in$${\cal{S}}^\prime$ and
$(f \star T)$$\in$${\cal{S}}^\prime$ while the unit $\mathbb{I}$$\in$${\cal{S}}^\prime$ plays the
role of the neutral element for the $\star$-product \cite{Gracia-Bondia:1987kw, Varilly:1988jk}. Now, let ${\cal{L}}$, ${\cal{R}}$ and ${\cal{M}}$ denote the following subspaces of ${\cal{S}}^\prime$
\begin{align}
{\cal{L}}=\big\{ T\in{\cal{S}}^\prime~:~~ T\star f\in{\cal{S}}, ~
\forall f\in {\cal{S}}\big\}, \quad 
{\cal{R}}=\big\{ T\in{\cal{S}}^\prime ~:~~ f\star T\in{\cal{S}},
~\forall f\in {\cal{S}}\big\}. \label{eq:left-right}
\end{align}
\begin{align}
{\cal{M}}={\cal{L}}\cap{\cal{R}}.\label{eq:Moyal-alg}
\end{align}
The algebra ${\cal{M}} $ defines the Moyal algebra, an
associative, unital, involutive $\star$-algebra. It
is a very large algebra involving the plane waves $e^{ipx}$, the
$\delta$ Dirac distribution and the ``coordinate'' functions
$x_\mu$ satisfying
\begin{align}
[x_\mu,x_\nu]_\star =i\Theta_{\mu\nu}
\end{align}
where $[a,b]_\star\equiv a\star b-b\star a$. Notice that this
relation, well defined on ${\cal{M}}$ but {\it{not}} on
${\cal{S}}$, may be used, and is
actually often used in the literature, as a convenient substitute for
a definition of the Moyal algebra. Useful properties of the $\star$-product valid on
${\cal{M}}$ are
\begin{subequations}
\begin{align}
\partial_\mu(f\star h)=\partial_\mu f\star h+f\star\partial_\mu
h,\qquad
(f\star h)^\dag=h^\dag\star f^\dag,\qquad
[x_\mu,f]_\star=i\Theta_{\mu\nu}\partial_\nu f, \label{eq:relat1}
\\
x_\mu\star f=(x_\mu.f)+{{i}\over{2}}\Theta_{\mu\nu}\partial_\nu f,
\qquad
x_\mu(f\star h)=(x_\mu.f)\star h
-{{i}\over{2}}\Theta_{\mu\nu}f\star\partial_\nu h ,\label{eq:relat2}
\end{align}
\end{subequations}
for any $f,h\in{\cal{M}}$. One of the most salient properties of
the $\star$-product is its non-locality which is obvious
from \eqref{eq:moyal}. To arrive at \eqref{eq:Moyal-alg}, one has to extend the $\star$-product on
${\cal{S}}^\prime\times{\cal{L}}$ and ${\cal{R}}\times
{\cal{S}}^\prime$ through
$\langle T\star T_L,f\rangle$$=$$\langle T,T_L\star f\rangle$, 
$\langle T_R\star T,f\rangle$$=$$ \langle T,f\star
T_R\rangle$, $\forall T$$\in$${\cal{S}}^\prime$, $\forall
T$$\in$${\cal{S}}^\prime$, $\forall
T_L$$\in{\cal{L}}$, $\forall
T_R$$\in$${\cal{R}}$, $\forall f$$\in$${\cal{S}}$ 
so that $(T\star T_L)$$\in$${\cal{S}}^\prime$ and $(T_R\star
T)$$\in$${\cal{S}}^\prime$. Then, one can show that for any 
$(T_1,T_2)$$\in$${\cal{M}}$, $(T_1\star
T_2)$$\in$${\cal{M}}$. The associativity of the $\star$-product extends easily on
${\cal{M}}$ as a mere consequence of the definition of the Moyal
algebra together with the definition of the duality bracket. 
It is easy to realize that ${\cal{M}}$
has a natural involution inherited from the complex conjugation
operation acting on ${\cal{S}}$ combined with the usual relation
$\langle T^\dag,f\rangle$$=$$\langle T,f^\dag\rangle^\star$,
$\forall T$$\in$${\cal{S}}^\prime$, $\forall f$$\in$${\cal{S}}$. In
particular, the relation $(T_1\star T_2)^\dag$=$T_2^\dag\star
T_1^\dag$ can be extended on ${\cal{M}}$.\par
\par
\section{Matter fields, connections, gauge potentials}
Before presenting the main steps of the computation, it is useful to clarify the mathematical status of
the various ingredients that will be involved, that is, matter
fields, connections, gauge potentials, curvatures. For a larger
amount of mathematical details, see e.g \cite{CONNES,CM},
\cite{Dubois-Violette:1989vq, Masson:1999}. The
framework presented below corresponds to the simplest option where
the relevant module is chosen to be the algebra
itself. It is also the most
widely studied situation in the literature, giving rise to the so
called "noncommutative field theories on flat Moyal spaces". This
has been partly motivated because a similar noncommutative structure
was suggested to emerge possibly from some limiting regimes of
string theory and matrix theory in magnetic backgrounds
\cite{Seiberg:1999vs, Witten:1985cc}.\par

\subsection{The free module case}
In the following, the Moyal algebra ${\cal{M}}$ is assumed to be endowed with a differential calculus
based on the derivations $\partial_\mu$. The usual concept of connections defined on
vector bundles in ordinary geometry can be consistently generalised
in noncommutative geometry to connections on (projective) modules
over an associative algebra. This is the route that will be followed
here. Let ${\cal{H}}$ be a right ${\cal{M}}$-module, equipped
with a Hermitian structure $h$, i.e a sesquilinear map $ h:{\cal{H}}\times{\cal{H}} \to {\cal{M}}$ verifying
\begin{align}
h(m_1\star f_1,m_2\star f_2) = f_1^\dag\star h(m_1,m_2)\star
f_2,\quad\forall f_1,f_2 \in {\cal{M}},\quad\forall m_1,m_2 \in
{\cal{H}} \label{eq:sesquilin}
\end{align}
A connection is defined algebraically by a linear map $\nabla_\mu: {\cal{H}} \to {\cal{H}}$
from ${\cal{H}}$ to ${\cal{H}}$ verifying:
\begin{align}
\nabla_\mu(m\star f)=\nabla_\mu(m)\star f+m\star\partial_\mu f,\
\forall m\in{\cal{H}},\quad \forall f\in{\cal{M}}\label{eq:nabla}
\end{align}
The connection is further assumed to preserve the hermitian
structure, that is
\begin{align}
\partial_\mu h(m_1,m_2)=h(\nabla_\mu m_1,m_2)+h(m_1,\nabla_\mu
m_2),\quad \forall m_1,m_2\in{\mathcal{H}}. \label{eq:hermite}
\end{align}
Up to now, the module has not been specified except it has been
assumed for definiteness to be a right ${\cal{M}}$-module. The
choice giving rise to the above mentioned noncommutative field
theories corresponds to ${\cal{H}} = {\cal{M}}$, that is, the
algebra plays the role of the module. This is the most simple
choice for ${\cal{H}}$ which thus becomes a free module. The
corresponding extension of the present material to 
the case ${\cal{H}} = {\cal{M}}^{\times N}$ is straightforward.\par

The choice ${\cal{H}} = {\cal{M}}$ has an immediate consequence.
Since the relevant module involves now the unit, it follows from
\eqref{eq:nabla} that the connection is entirely determined by
its action $\nabla_\mu(\mathbb{I})$ on the unit $\mathbb{I} \in
\mathcal{M}$, denoted by
\begin{align}
\nabla_\mu^A(\mathbb{I})\equiv-iA_\mu,\label{eq:connection-def}
\end{align}
Indeed, by setting $m$$=$$\mathbb{I}$ in \eqref{eq:nabla}, one
readily obtains
\begin{align}
\nabla^A_\mu(\mathbb{I}\star f) = \nabla^A_\mu(\mathbb{I})\star
f+\partial_\mu f \equiv
\partial_\mu f-iA_\mu\star f
\end{align}
This therefore can serve as defining a noncommutative analog of the
gauge potential $A_\mu$ in ${\cal{M}}$. Furthermore, it can be
realized that for ${\cal{H}} = {\cal{M}}$, a hermitian structure is
provided by
\begin{align}
h(f_1,f_2) = f_1^\dag\star f_2 \label{eq:complexstruct}
\end{align}
which ensures that the above connections are hermitian
connections provided $A_\mu^\dag$$=$$A_\mu$.\par

\subsection{Gauge transformations}
First recall that when ${\cal{H}} \ne {\cal{M}}$, a morphism of
module, says $\gamma$, satisfies by definition
\begin{align}
\gamma(m\star f) =
    \gamma(m)\star f,\quad\forall m \in {\cal{H}},\quad\forall f \in
    {\cal{M}}
\end{align}
In the present situation, where ${\cal{H}}={\cal{M}}$ is assumed, the gauge
transformations hereafter denoted by $\gamma$ are determined by the
automorphisms of the module ${\cal{M}}$ preserving the
hermitian structure $h$: $\gamma \in Aut_h({\cal
{M}})${\footnote{keeping in mind that ${\cal{M}}={\cal{H}}$ is
considered as a hermitian module over itself}}. Namely, the gauge
transformations must satisfy
\begin{subequations}
\begin{align}
\gamma(f) &=\gamma(\mathbb{I}\star f)=\gamma(\mathbb{I})\star f\
,\quad \forall f\in{\mathcal{M}} \label{eq:unitary2}
\\
h\big(\gamma(f_1),\gamma(f_2)\big) &= h(f_1,f_2) \quad \forall
f_1,f_2 \in \mathcal{M} \qquad , \label{eq:unitary}
\end{align}
\end{subequations}
It can be verified that the combination of \eqref{eq:unitary} with
\eqref{eq:complexstruct} applied to $\gamma(\mathbb{I})$ implies
automatically
\begin{align}
\gamma(\mathbb{I})^\dag \star \gamma(\mathbb{I}) = \mathbb{I}
\label{eq:gege-1}
\end{align}
From this, it follows that the gauge transformations are
entirely determined by $\gamma(\mathbb{I})\in {\cal{U}}(\cal{M})$,
where ${\cal{U}}(\cal{M})$ is the group of unitary elements of
$\cal{M}$. From now on, we set
\begin{align}
\gamma(\mathbb{I}) \equiv g
\end{align}
According to \eqref{eq:unitary2}, the action of the gauge group
as given above on any matter field $\phi \in {\cal{M}}$ can be
defined by
\begin{equation}
\phi^g=g\star\phi\label{eq:fundam}
\end{equation}
for any $g \in {\cal{U}}(\cal{M})$. This simply
expresses the fact that the gauge transformation is a morphism of
module. In more physical words, this may be viewed as the
noncommutative analogue of the transformation of the matter fields
under the ``fundamental representation of the gauge group''. Note
that one has obviously $g^\dag\star g = g\star g^\dag = \mathbb{I}$
in view of \eqref{eq:gege-1}.\par

The action of ${\cal{U}}({\cal {M}})$ on the connection
$\nabla^A_\mu$ is defined by
\begin{align}
(\nabla^A_\mu)^\gamma(\phi)
=\gamma(\nabla^A_\mu(\gamma^{-1}\phi)),\quad \forall
\phi\in{\cal{M}}. \label{eq:transjauge}
\end{align}
By further using $\gamma(\phi) = \gamma(\mathbb{I} \star \phi) = g
\star \phi$ together with the expression of the covariant derivative
\begin{align}
\nabla_\mu^A(\phi)=\partial_\mu\phi-iA_\mu\star\phi
\label{eq:covder}
\end{align}
and the fact that one can define
\begin{align}
(\nabla_\mu^A)^g \equiv\partial_\mu-iA^g_\mu
\end{align}
one obtains easily the following gauge transformation for the gauge
potential $A_\mu$
\begin{align}
A_\mu^g=g\star A_\mu\star g^\dag+ig\star\partial_\mu g^\dag .
\label{eq:gaugetransf}
\end{align}

It appears that another type of transformations given by 
\begin{align}
\phi^U=U\star \phi\star U^\dag\equiv\alpha(\phi) \label{eq:adjoint-de}
\end{align}
for any $U\in{\cal{U}}({\cal{M}})$ has been also considered in the literature. In physical words, this transformation could be viewed as a noncommutative analog of the so called "gauge transformation in the adjoint representation". It is instructive to confront the actual mathematical status of these transformations to the algebraic framework developed above. In fact, \eqref{eq:adjoint-de} defines an automorphism $\alpha$ of algebra, since one has obviously
\begin{align}
\alpha(\phi_1\star\phi_2)=\alpha(\phi_1)\star\alpha(\phi_2) \label{eq:automalg1}
\end{align}
However, this transformation does not correspond to an automorphism of module (which should satisfy $\alpha(\phi_1\star\phi_2)$$=$$\alpha(\phi_1)\star \phi_2$) except when $U$ is in the center of ${\cal{M}}$ which in the present case is equal to ${\mathbb{C}}$. Within the present framework, the transformations \eqref{eq:adjoint-de} cannot be viewed, strictly speaking, as gauge transformations which, by their very definition, must be related to morphisms of module. It would be temptating to introduce a "covariant derivative" whose form suggested by \eqref{eq:adjoint-de} is given by
\begin{align}
D_\mu(\phi)=\partial_\mu\phi-i[A_\mu,\phi]_\star \label{eq:derivationcov}
\end{align}
Covariance under \eqref{eq:adjoint-de} $(D_\mu(\phi))^U$$=$$U\star (D_\mu(\phi))\star U^\dag$ is insured provided
\begin{align}
A_\mu^U=U\star A_\mu\star U^\dag+iU\star\partial_\mu U^\dag 
\end{align}
It is easy to realize that $D_\mu$ is a derivation of the Moyal algebra ${\cal{M}}$ since it satisfies a Leibnitz rule
\begin{align}
D_\mu(\phi_1\star\phi_2)=D_\mu(\phi_1)\star\phi_2+\phi_1\star D_\mu(\phi_2) 
\end{align}
which however does not coincides with \eqref{eq:nabla} defining a noncommutative connection. However, the above noncommutative analog of "gauge transformation in the adjoint representation" can be understood in terms of actual noncommutative gauge transformations provided the initial algebra ${\cal{M}}$ is enlarged to $\mathcal{M}$ by $\mathcal{M} \otimes \mathcal{M}^o$, where $\mathcal{M}^o$ is the opposite algebra as shown for instance in \cite{CM} which amounts to deal with real structure instead of hermitian structure. This will not be further considered in this talk.\par

\subsection{Canonical gauge-invariant connection}
In the present Moyal framework, it turns out that the following relation holds
\begin{align}
\partial_\mu\phi=[i\xi_\mu,\phi]_\star, \quad \xi_\mu\equiv-\Theta^{-1}_{\mu\nu}x_\nu \label{eq:inner-der}
\end{align}
simply expressing the fact that the derivative $\partial_\mu$ in
$\cal{M}$ is an inner derivative. In fact, all derivatives in the Moyal algebra are inner derivatives.\par
As already pointed out during this conference \cite{Massonconf}, the existence of inner derivations \eqref{eq:inner-der} implies necessarily the existence of a "canonical connection" that remains invariant under the gauge transformations. Canonical gauge-invariant connections are not new in
noncommutative geometry as already mentioned in earlier
studies \cite{Dubois-Violette:1989vq}. Here, the gauge-invariant connection is defined by $\xi_\mu$. Indeed, the connection
$\nabla_\mu^\xi$ verifies
\begin{align}
\nabla^\xi_\mu\phi=\partial_\mu\phi-i\xi_\mu\star\phi=-i\phi\star\xi_\mu
\end{align}
where the second equality stems simply from \eqref{eq:inner-der}. Then, it is easy to realise that
\begin{align}
(\nabla_\mu^\xi)^g(\phi)= g\star(\nabla_\mu^\xi(g^\dag\star\phi))
=-i\phi\star\xi_\mu=\nabla_\mu^\xi\phi
\end{align}
which shows that $\nabla_\mu^\xi$ is invariant under the gauge
transformations, leading to
\begin{align}
  \xi_\mu^g=\xi_\mu \label{eq:invar-xi}
\end{align}
This could have been checked directly by combining the actual
expression for $\xi_\mu$ with \eqref{eq:gaugetransf} and \eqref{eq:inner-der}. \par
In the present case, the existence of the gauge-invariant connection defined above singles out a natural tensorial form  defined as
\begin{align}
\nabla^A_\mu-\nabla^\xi_\mu=-i(A_\mu-\xi_\mu)\equiv-i\mathcal{A}_\mu
\label{eq:arondvrai}
\end{align}
This tensorial form coincides with the so called covariant coordinates (see e.g.\ \cite{Douglas:2001ba}) that are sometimes introduced in the context of string theories. \par
Given a connection $\nabla_\mu^A$, the corresponding curvature is
\begin{align}
F_{\mu\nu}^A\equiv i[\nabla^A_\mu,\nabla^A_\nu]_\star=&
\partial_\mu A_\nu-\partial_\nu A_\mu-i[A_\mu,A_\nu]_\star\nonumber\\
=&\Theta_{\mu\nu}^{-1}
-i[\mathcal{A}_\mu,\mathcal{A}_\nu]_\star=F_{\mu\nu}^\xi-i[\mathcal{A}_\mu,\mathcal{A}_\nu]_\star
\label{eq:curvature}
\end{align}
where the two rightmost equalities stem from \eqref{eq:inner-der} and
\eqref{eq:arondvrai}. The gauge transformations for ${\cal{A}}_\mu$ and $F_{\mu\nu}^A$ are readily obtained and are given by
\begin{align}
\mathcal{A}_\mu^g=g\star\mathcal{A}_\mu\star g^\dag,\quad (F_{\mu\nu}^A)^g=g\star F_{\mu\nu}^A\star g^\dag \label{eq:jaugetransf}
\end{align}
Note that the invariant connection defined by $\xi_\mu$ is a
constant curvature connection since one has
\begin{align}
F_{\mu\nu}^\xi =
\Theta_{\mu\nu}^{-1}
\end{align}
\par
\subsection{The minimal coupling prescription.}

In the rest of this talk, I will assume that the action of the gauge group on the matter fields
$\phi$ is given by \eqref{eq:fundam}. Owing to the special role
played by the coordinate functions $x_\mu$ through the invariant
gauge potential $\xi_\mu$ involved in $\nabla^\xi_\mu$ and \eqref{eq:inner-der}, a natural choice for the minimal coupling prescription \cite{PAPER1} is obtained by performing the usual substitution
\begin{align}
\partial_\mu\to\nabla_\mu^A
\end{align}
in the considered action {\it{provided this latter is
reexpressed in terms of}} $\partial_\mu$ {\it{and}}
$\nabla_\mu^\xi$, using in particular the following identity:
\begin{align}
\widetilde{x}_\mu\phi=\widetilde{x}_\mu\star\phi-i\partial_\mu\phi
=-i(\partial_\mu\phi-2i\xi_\mu\star\phi)
= -2i\nabla_\mu^\xi\phi+i\partial_\mu\phi. \label{eq:identitycouplage}
\end{align}
Using \eqref{eq:identitycouplage}, the
minimal coupling prescription can be conveniently written as
\begin{align}
\partial_\mu\phi&\mapsto \nabla_\mu^A\phi=
\partial_\mu\phi-iA_\mu\star\phi, \label{eq:coup1}\\
\widetilde{x}_\mu\phi &\mapsto
-2i\nabla_\mu^\xi\phi+i\nabla_\mu^A\phi=\widetilde{x}_\mu\phi+
A_\mu\star\phi.\label{eq:coup2}
\end{align}
Note that gauge invariance of the resulting action functional is
obviously obtained thanks to the relation
\begin{align}
(\nabla^{A,\xi}_\mu(\phi))^g = g\star(\nabla^{A,\xi}_\mu(\phi))
\end{align}
This prescription will be used in a while to couple the relevant matter action with harmonic term to an external gauge potential.
\section{Renormalisable actions for matter}
\subsection{The harmonic solution}
Up to now, I have recalled the main features of the Moyal algebra ${\cal{M}}$ within the relevant algebraic framework of noncommutative geometry. I have also assumed ${\cal{M}}$ plays the role of the module so that we work with a simple free module from which are defined the fields and connections to be further used in some actions. Let us briefly review some salient features of the present situation about renormalisable actions for matter. \par
As far as I know, no firm guiding principle ruling the construction of actions on ${\cal{M}}$ has been identified so far. In $D$$=$$4$, a rather natural noncommutative analog of the commutative Yang-Mills action stemming from the Connes-Lott action functional \cite{VGAYRAL}, the spectral action principle \cite{VASSILEVICH} is given by
\begin{align}
S_{YM}=\int d^4x F_{\mu\nu}\star F_{\mu\nu} \label{eq:yangnaif}
\end{align}
where $F_{\mu\nu}$ is defined by \eqref{eq:curvature}. This however cannot be directly extended to the matter actions. Nevertheless, the expression \eqref{eq:yangnaif} suggested in the old days the following simple recipe: Naive noncommutative candidates for matter actions on ${\cal{M}}$ may be obtained by replacing all the ordinary local of their commutative counterparts by non-local $\star$-products. This procedure preserves the locality of the kinetic term since one has $\int d^4x\partial_\mu\phi^\dag\star\partial_\mu\phi$$=$$\int d^4x\partial_\mu\phi^\dag\partial_\mu\phi$
thanks to \eqref{eq:tracial}. But it turns any commutative local polynomial interaction into a non local interaction, as a consequence of the definition of the $\star$-product. For instance, one has
\begin{align}
    \int d^4x(\phi^\dag\star\phi\star\phi^\dag\star\phi)(x)
&=\frac{1}{\pi^4\theta^4}\int
    \prod_{i=1}^4d^4x_i\,\phi^\dag(x_1)\phi(x_2)\phi^\dag(x_3)\phi(x_4) \nonumber
    \end{align}
\begin{align}
 \times\delta(x_1-x_2+x_3-x_4)e^{-i\sum_{i<j}(-1)^{i+j+1}x_i\wedge
   x_j }. \nonumber
 \end{align}
where $x\wedge y$$\equiv$$2x_\mu\Theta^{-1}_{\mu\nu}y_\nu$.
The non locality of the interaction is responsible for the possible occurrencee of the UV/IR-mixing
\cite{Minwalla:1999px,Chepelev:1999tt}, that makes the
renormalisability very unlikely. It results from the existence of dangerous non-planar {\footnote{The terminology ``planar'' and ``non planar'' diagrams can be most easily understood within the matrix base formalism where diagrams become ribbon diagrams. Planar (resp. non planar) diagrams can (resp. cannot) be drawn on a plan are called planar (resp. non planar).}}diagrams which, albeit UV finite, become singular at exceptional (low) external momenta. This triggers the appearancee of UV divergences in higher
order diagrams in which they are involved as subdiagrams,signallingg that UV and IR scales are related in a non-trivial way which should in principle invalidate a Wilson-type renormalisation scheme \cite{Wilson:1973jj,Polchinski:1983gv}. Consider for instance the action for the real-valued scalar $\varphi^4$ theory in $D$$=$$4$:
\begin{align}
S_{0\varphi}=\int d^4x \big(\frac{1}{2}\partial_\mu\varphi\star\partial_\mu\varphi+\frac{m^2}{2}\varphi\star\varphi+\frac{g}{4}\varphi\star\varphi\star\varphi\star\varphi \big)\label{eq:varphiact}
\end{align}

The one-loop tadpole part of the two-point function involves two contributions. The planar diagram contribution is $C_p$$\sim$$ g\int \frac{d^4k}{(2\pi)^4}\frac{1}{k^2+m^2}$
and exhibits the expected UV divergence in four dimensions. However, the non planar contribution is
\begin{align}
C_{np}=\frac{g}{12} \int \frac{d^4k}{(2\pi)^4}\frac{e^{ik\wedge p}}{k^2+m^2}=\frac{gm}{48\pi^2}{\sqrt{(\tilde p)^2}}K_1(m{\sqrt{(\tilde p)^2}})
\end{align}
where $\tilde p$$\equiv$$\Theta_{\mu\nu}p_\nu$, the exponential oscillating factor in the integrand comes simply from the non locality of the vertex and $K_1$ denotes the modified Bessel function of rank 1 ($p$ is the external momentum). Although this latter contribution is UV finite, it exhibits an infrared singularity, since one has
$C_{np}$$\sim$$ \frac{1}{(\tilde p)^2}$ for $p\to 0$
which holds thanks to $K_1(x)$$\sim$$x^{-1}$ for $x$$\to$$0$. This implies that, if $\Lambda$ denotes some UV cut-off used to regularise the ultraviolet behaviour of the integrals, the two limits $\Lambda$$\to$$\infty$ and $p$$\to$$0$ do not commute. This is, in essence, the UV/IR mixing phenomenon first exhibited in \cite{Minwalla:1999px}. Moreover, such a tadpole triggers theappearancee of UV divergence in higher order diagrams in which it is involved as a subdiagram. This divergence is non local and therefore cannot be absorbed into a mass redefinition so that the scalar model \eqref{eq:varphiact} is not a priori renormalisable. A somehow similar conclusion applies to the naive noncommutative extension of the Yang-Mills action \eqref{eq:yangnaif} indicating clearly that some principle(s) should supplement or replace the initial naive recipe. As far as I know, two possible solutions have been identified so far, each having its own merits and drawbacks. In the rest of this talk, I will focus only on the "harmonic solution" proposed by Grosse and Wulkenhaar \cite{Grosse:2004yu}. The second possible solution is related to the so called twisted noncommutative field theories (for more details see e.g \cite{RSZABOBIS} and references therein).\par

In \cite{Grosse:2004yu}, Grosse and Wulkenhaar showed that the UV/IR-mixing of \eqref{eq:varphiact} can be suppressed by supplementing the initial action with a harmonic oscillator term
leading to a renormalisable noncommutative quantum field theory. This is the ``harmonic solution''. The modified action is ($\widetilde{x}_\mu$$=$$2\Theta^{-1}_{\mu\nu}x_\nu$)
\begin{align}
S_{\varphi}=\int d^4x \big(\frac{1}{2}\partial_\mu\varphi\star\partial_\mu\varphi+\frac{\Omega^2}{2}(\widetilde{x}_\mu\varphi)\star(\widetilde{x}_\mu\varphi)+\frac{m^2}{2}\varphi\star\varphi+\frac{g}{4}\varphi\star\varphi\star\varphi\star\varphi \big)\label{eq:varphimod}
\end{align}
where $\Omega$$\in$$[0,$ $1]$ for reasons given below. The actual interpretation of the harmonic term is so far not quite clear. The initial beautiful proof for the renormalisability of \eqref{eq:varphimod} has been carried out within the matrix-base formalism. Basically, using the fields expansion in the matrix-base, one maps \eqref{eq:varphimod} onto an infinite-dimensional complex-valued matrix model. The technical key point is the computation of the propagator as the formal inverse of an infinite dimensional matrix in terms of the Meixner polynomials. By further combining the propagators to the renormalisation group machinery with sharp bounds put on the matrix indices and using the Wilson-Polchinski equations, one can then prove \cite{Grosse:2004yu} that the modified action \eqref{eq:varphimod} is renormalisable to all orders. The introduction of the harmonic term produces in some sense a kind of infrared cut-off in the theory and allows the UV and IR scales to decouple. It turns out that \eqref{eq:varphimod} is covariant under a duality transformation \cite{Langmann:2002cc}, sometimes called the ``Langman-Szabo'' duality (LS duality). Roughly speaking, this expresses the fact that the form of \eqref{eq:varphimod} is, up to rescaling factors, not altered through $\widetilde x_\mu$$\leftrightarrows$$\partial_\mu$. The LS duality is defined{\footnote{${\hat{X}}$ is the cyclic Fourier transform of $X$ given by ${\widehat{f}}(p_i)$$=$$\int d^4x e^{i(-1)^ip_i\cdot x_i}f(x_i)$; the coordinates appearing in cyclic order in the quartic interaction are labelled as $x_i$, $i$$=$$1,2,3,4$ and similarly for the momenta $p_i$ }} by
\begin{align}
{\widehat{\varphi}}(p)\leftrightarrows\pi^2{\sqrt{|det\Theta|}}\varphi(x), \quad p_\mu\leftrightarrows \widetilde x_\mu \label{eq:LSD}
\end{align}
which applied on the action yields
\begin{align}
S_\varphi(\varphi;m,g,\Omega) \leftrightarrows \Omega^2S_\varphi(\varphi;\frac{m}{\Omega},
\frac{g}{\Omega^2},\frac{1}{\Omega})
\end{align}
so that $\Omega$ can be consistently restricted to $\Omega$$\in$$[0,$ $1]$. Note that \eqref{eq:varphimod} is invariant for $\Omega$$=$$1$. So far, the requirement of LS duality has been helpfull to construct other renormalisable actions for complex-valued scalar fields. However, more investigations are needed to clarify the actual role of the L-S duality \cite{Langmann:2002cc} in the control of the UV/IR-mixing and renormalisability. While it is obvious that the implementation of LS-duality can be easily achieved within matter actions with polynomial interaction terms, the situation changes drastically for Yang-Mills type theories as it will be shown in the next section. Let us now recall the main features of most of the presently known renormalisable field theories defined on Moyal spaces. Note
that interesting solvable noncommutative scalar field theories have been considered in
\cite{Grosse:2005ig,Grosse:2006qv,Grosse:2006tc} to which I refer for more details. \par
\subsection{The  4-dimensional ``harmonic'' complex scalar model}
The corresponding classical action , covariant under the LS-duality, is 
\begin{equation}
S(\phi)=\int d^4x\big(\partial_\mu\phi^\dag\star\partial_\mu\phi
+\Omega^2(\widetilde{x}_\mu\phi)^\dag\star(\widetilde{x}_\mu\phi)
+m^2\phi^\dag\star\phi\big)(x)+S_{int},\label{eq:actionharm}
\end{equation}
$\phi$ is a complex scalar field with mass $m$, the parameters $\Omega$ and $\lambda$
are dimensionless and the term involving $\Omega$ is the (complex-valued) scalar 
counterpart of the harmonic oscillator term. The interaction term $S_{int}$ is
\begin{equation}
S_{int}=S_{int}^0+S_{int}^{NO}
=\int \lambda(\phi^\dag\star\phi\star\phi^\dag\star\phi)(x)
+\kappa(\phi^\dag\star\phi^\dag\star\phi\star\phi)(x).\label{eq:scalarint}
\end{equation}
It turns out that \eqref{eq:actionharm} restricted to $S_{int}^O$ ($\kappa$$=$$0$) is renormalisable for any value of $\Omega$ \cite{Gurau:2005gd} (see also \cite{Vignes-Tourneret:2006nb}). The effect of the inclusion of $S_{int}^{NO}$ on the renormalisability is not known. However in that latter case, it can be shown \cite{NOUS1} that a term $\sim$ $ \kappa\phi^\dag{\widetilde{x}}_\mu\partial_\mu\phi$
is induced at the one-loop order whose coefficient is such that it ``completes the square in the quadratic part'': the quadratic part of the action can then be cast into the form:$\sim$$|\partial_\mu\phi$$+$ $i\alpha{\widetilde{x}}_\mu\phi|^2$ where 
$\alpha$ is some constant. This can be explicitely verified by a direct computation in the $x$-space, using the 
Mehler expression for the propagator $C(x,y) \equiv \langle\phi(x)\phi^\dag(y)\rangle$ obtained by solving
$(\Delta_x+{\widetilde{\Omega}}^2x^2+m^2)C(x,y) =\delta(x - y)$ which gives (${\widetilde{\Omega}}\equiv
2{{\Omega}\over{\theta}}$)
\begin{equation}
C(x,y)={{\Omega^2}\over{\pi^2\theta^2}}\int_0^\infty \!\!
{{dt}\over{\sinh^2(2{\widetilde{\Omega}}t)}}\exp\Big(
-{{{\widetilde{\Omega}}}\over{4}}\coth({\widetilde{\Omega}}t)(x{-}y)^2
-{{{\widetilde{\Omega}}}\over{4}}\tanh({\widetilde{\Omega}}t)(x{+}y)^2-m^2t\Big),
\label{eq:propag}
\end{equation}
\subsection{The LSZ-type models}
The corresponding action is given by
\begin{align}
S_{LSZ}(\phi)=\int d^4x\big((\partial_\mu\phi
+i\Omega{\widetilde{x}}_\mu\phi)^\dag\star
(\partial_\mu\phi+i\Omega{\widetilde{x}}_\mu\phi)
+m^2\phi^\dag\star\phi\big)(x)+S_{int}\label{eq:actionlsz}
\end{align}
where $S_{int}$ has been given above. For the initial formulation of the LSZ model, see \cite{Langmann:2003cg}. The LSZ model has been shown to be renormalisable to all orders whenever $\kappa$$=$$0$. The proof, as sketched in \cite{Gurau:2005gd}, is somehow similar to the
one given in \cite{Vignes-Tourneret:2006nb} for the noncommutative
Gross-Neveu model. At the present time, the actual impact of
interaction terms as given by $S_{int}^{NO}$ on the renormalisability
of the above models is not known. However, it is likely that the LSZ model involving both $S_{int}^{O}$ and $S_{int}^{NO}$ is renormalisable to all order.\par
At this level, it is
instructive to interpret the action \eqref{eq:actionlsz} in the light
of the algebraic framework that has been developed. In fact, the operator
$\partial_\mu+i\Omega{\widetilde{x}}_\mu$ is actually related to a
connection $\nabla^{\zeta}_\mu$ with
\begin{align}
\zeta_\mu=\frac{2\Omega}{1+\Omega}\xi_\mu ,\label{eq:zeta}
\end{align}
since the following relation
\begin{align}
(\partial_\mu+i\Omega{\widetilde{x}}_\mu)\phi
=(1+\Omega)\Big(\partial_\mu\phi
-i{{2\Omega}\over{1+\Omega}}\xi_\mu\star\phi\Big)
=(1+\Omega)\nabla^{\zeta}_\mu(\phi)
\end{align}
holds in view of \eqref{eq:covder}. The action \eqref{eq:actionlsz}
can then be rewritten as
\begin{align}
S_{LSZ}(\phi)=\int d^4x\big((1+\Omega)^2
(\nabla^{\zeta}_\mu(\phi))^\dag\star\nabla^{\zeta}_\mu(\phi)
+m^2\phi^\dag\star\phi\big)(x)+S_{int},
\end{align}
where $\zeta$ is given by \eqref{eq:zeta} which, for $\Omega \ne 0$,
makes explicit the invariance of the action under the gauge
transformations $\phi^g = g\star\phi$ for any
$g\in{\cal{U}}({\cal{M}})$. Notice that a similar comment applies to
the noncommutative version of the (two-dimensional) Gross-Neveu model
considered recently in \cite{Vignes-Tourneret:2006nb}, \cite{Lakhoua:2007ra} to which we turn now on.\par

\subsection{The 2-D orientable Gross-Neveu model}
The Moyal algebra can be trivially adapted to the two-dimensional case. The action for the non-commutative Gross-Neveu model first considered in \cite{Vignes-Tourneret:2006nb} can be written as
\begin{align}
S_{GN}=&\int 
    d^2x\big[{\bar{\psi}}(-i{\slashed{\partial}}+\Omega{\slashed{\widetilde{x}}}+m+\kappa\gamma_5)\psi-
    \sum_{A=1}^3\frac{g_A}{4}({\cal{J}}^{A}\star{\cal{J}}^{A})(x)\big]\label{eq:actiongneveu}
 \end{align}
 with
 \begin{align}
    {\cal{J}}^{A}=&{\bar{\psi}}\star\Gamma^{A}\psi,\ 
    \Gamma_{1}=\mathbb{I},\,\Gamma_{2}=\gamma^{\mu},\,\Gamma_{3}=\gamma_{5} \label{eq:courants}  
    \end{align}
where $\slashed{a}=a_\mu\gamma^\mu$ and a summation over the Lorentz indices $\mu$ is understood in the interaction term involving ${\cal{J}}_2$,  ${\cal{J}}_2\star{\cal{J}}_2$$=$$\sum_\mu({\bar{\psi}}\star\gamma_\mu\psi)\star({\bar{\psi}}\star\gamma^\mu\psi)$. The Clifford algebra for the 2D anti-Hermitian gamma matrices satisfy $\{\gamma^{\mu},\gamma^{\nu}\}=-2\delta^{\mu\nu}$ and $\gamma_5=\imath\gamma^{0}\gamma^{1}$. Here, the field $\psi$ denotes a $2N$-component spinor field where $N$ is the number of colors. In \eqref{eq:actiongneveu}, the term $\Omega{\bar{\psi}}{\slashed{\widetilde{x}}}\psi$is the Fermionic counterpart of the harmonic oscillator term. The minus sign affecting the four-Fermion interaction term in \eqref{eq:actiongneveu} is a meregeneralisationn of the interaction term in the commutative Gross Neveu model for which asymptotic freedom is obtained. Six independent four-fermion interactions can in principle be constructed. The three interaction terms in  which $\psi$ and ${\bar{\psi}}$ alternate, namely $\sum_{a,b}{\bar{\psi}}_a\star\psi_b\star{\bar{\psi}}_a\star\psi_b$, $\sum_{a,b}{\bar{\psi}}_a\star\psi_b\star{\bar{\psi}}_b\star\psi_a$, $\sum_{a,b}{\bar{\psi}}_a\star\psi_a\star{\bar{\psi}}_b\star\psi_b$ (the sum runs over color indices $a$, $b$), gives rise after suitable Fierz transformations to the interaction term in \eqref{eq:actiongneveu}. This action has been shown to be renormalisable to all orders in \cite{Vignes-Tourneret:2006nb}. Three other interaction terms with adjacent $\psi$ and ${\bar{\psi}}$ could in principle be written, namely $\sum_{a,b}{\bar{\psi}}_a\star{\bar{\psi}}_b\star\psi_a\star\psi_b$, $\sum_{a,b}{\bar{\psi}}_a\star{\bar{\psi}}_b\star\psi_b\star\psi_a$, $\sum_{a,b}{\bar{\psi}}_a\star{\bar{\psi}}_a\star\psi_b\star\psi_b$. Although not explicitely proven at the present time, it seems likely that the Gross-Neveu model involving both type of interactions is renormalisable.\par 
It can be easily realised that $S_{GN}$ can be cast into the form
\begin{align}
S_{GN}=\int d^2x\big(-i(1+\Omega){\bar{\psi}}\gamma^\mu
\nabla_\mu^\zeta\psi+m{\bar{\psi}}\psi \big)(x)+...,\label{eq:GN}
\end{align}
where the ellipses denote interaction terms. Recall that the same covariant derivative already appeared in the LSZ-type models. In physical words, these two latter actions can be interpreted as matter actions already
coupled to an external (background) gauge potential $\zeta_\mu$ (while
the action \eqref{eq:actionharm} does not obviously support this
interpretation).\par

\section{The one-loop effective gauge action.}
Let us start from the action $S(\phi,A)$ obtained byminimallyy coupling \eqref{eq:actionharm} to an external gauge potential through the minimal coupling prescription introduced above. Indeed, combining \eqref{eq:coup1}, \eqref{eq:coup2} with \eqref{eq:actionharm}, one easily obtains the gauge-invariant action \cite{PAPER1}
\begin{align}
S(\phi,A) =& S(\phi)+ \int d^4x \ \big((1+\Omega^2)\phi^\dag\star
(\widetilde{x}_\mu A_\mu)\star\phi\nonumber\\ 
&-(1-\Omega^2)\phi^\dag\star A_\mu \star\phi\star
\widetilde{x}_\mu+
(1+\Omega^2)\phi^\dag\star A_\mu\star
A_\mu\star \phi\big)(x),\label{eq:harmcoupled}
\end{align}
where $S(\phi)$ is given by \eqref{eq:actionharm} with $S_{int}$ restricted to its gauge-invariant part
$S_{int}^O$, see \eqref{eq:scalarint}. We are now in position to compute the effective action $\Gamma(A)$ obtained by integrating over the scalar field $\phi$ in $S(\phi,A)$, for any value of $\Omega \in [0,1]$. The computation can be conveniently carried out within the $x$-space formalism \cite{Gurau:2005gd}. For a similar computation performed in the matrix-base, see \cite{Grosse:2007hh}. There are several motivations to perform such a calculation. In particular, it can be expected that the expression of $\Gamma(A)$ should be a hepfull tool to guess possible form(s) for a candidate as a renormalisable gauge action. This would permits one to determine whether or not some additional terms beyond the expected $F_{\mu\nu}\star F_{\mu\nu}$ appear in the effective action, in what extend the initial harmonic term in \eqref{eq:harmcoupled} survive in $\Gamma(A)$ action and whether or not some relic of the Langmann-Szabo shows up in the effective action.\par
\begin{figure}[!htb]
  \centering
  \includegraphics[scale=1]{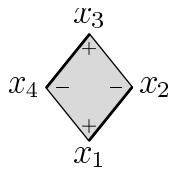}
  \caption[The Vertex]{\footnotesize{Graphical representation for the
      vertex in the $x$-space, obtained from \eqref{eq:interaction}.
      The plus-sign (resp.\ minus-sign) appearing in the rhombus
      corresponds to incoming (resp.\ outgoing) external line
      associated with ${\phi^\dag}$ (resp.\ $\phi$).}}
  \label{fig:vertex}
\end{figure}

Recall that the effective action is formally obtained from
\begin{align}
e^{-\Gamma(A)}\equiv \int D\phi D\phi^\dag e^{-S(\phi,A)}
=\int D\phi D\phi^\dag e^{-S(\phi)} e^{-S_{int}(\phi,A)}, \label{eq:defact}
\end{align}
where $S(\phi)$ is given by \eqref{eq:actionharm} and
$S_{int}(\phi,A)$ can be read off from \eqref{eq:harmcoupled} and
\eqref{eq:actionharm}. At the one-loop order, \eqref{eq:defact}
reduces to
\begin{align}
e^{-\Gamma_{1loop}(A)}=\int D\phi D\phi^\dag e^{-S_{free}(\phi)}
e^{-S_{int}(\phi,A)},
\end{align}
where $S_{free}(\phi)$ is simply the quadratic part of
\eqref{eq:actionharm}. The corresponding diagrams are depicted on the
figures \ref{fig:1-point}-\ref{fig:4-point}. These diagrams can be computed from the propagator and interaction
vertices derived from \eqref{eq:actionharm}. The scalar
propagator $C(x,y) \equiv \langle\phi(x)\phi^\dag(y)\rangle$ has been already given in \eqref{eq:propag}. The interaction vertices can in principle be read off from the RHS of
\begin{subequations}
  \label{eq:interaction}
  \begin{align}
    \int d^4x(\phi^\dag\star\phi\star\phi^\dag\star\phi)(x)
&=\frac{1}{\pi^4\theta^4}\int 
    \prod_{i=1}^4d^4x_i\,\phi^\dag(x_1)\phi(x_2)\phi^\dag(x_3)\phi(x_4)
\label{eq:InteractionDvpee}
\\
 &\times\delta(x_1-x_2+x_3-x_4)e^{-i\sum_{i<j}(-1)^{i+j+1}x_i\wedge
   x_j }. \nonumber
  \end{align}
\end{subequations}
The generic graphical representation of a vertex is depicted on the
figure~\ref{fig:vertex}. The non-locality of the interaction is
conveniently represented by the rhombus appearing on
fig.~\ref{fig:vertex} whose vertices correspond to the $x_i$'s
occurring in \eqref{eq:interaction}. It is useful to represent the
alternate signs in the delta function of \eqref{eq:interaction} by
plus- and minus-signs, as depicted on the figure. By convention, a
plus-sign (resp.\ minus-sign) corresponds to an incoming field
$\phi^\dag$ (resp.\ outgoing field $\phi$). This permits one to define
an orientation on the diagrams obtained from the loop expansion.\par
\begin{figure}[!htb]
  \centering
  \includegraphics[scale=1]{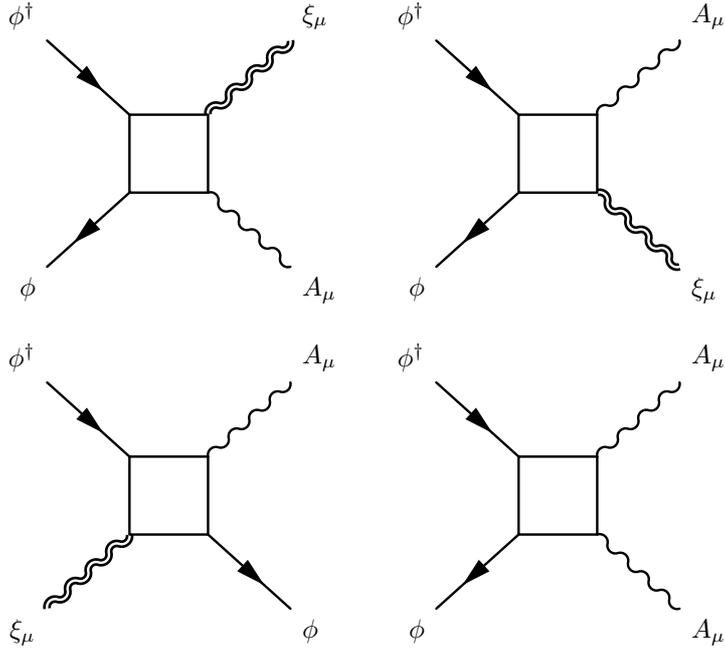}
  \caption[The vertices]{\footnotesize{Graphical representation for
      the vertices carrying the external gauge potential $A_\mu$
      involved in the action \eqref{eq:harmcoupled}. The overall
      factor affecting the two uppermost vertices is $(1 + \Omega^2)$.
      From left to right, the overall factors affecting the lower
      vertices are respectively equal to $- 2(1 - \Omega^2)$ and $- (1
      + \Omega^2)$.}}
  \label{fig:vertices}
\end{figure}

The additional vertices involving $A_\mu$ and/or $\xi_\mu$ and
generated by the minimal coupling that will be relevant for the present calculation can be obtained by combining
\eqref{eq:harmcoupled} with the generic
relation
\begin{align}
\int d^4x(f_1\star f_2\star f_3\star f_4)(x)=\frac{1}{\pi^4\theta^4}\int
\prod_{i=1}^4d^4x_i\, f_1(x_1) f_2(x_2) f_3(x_3) f_4(x_4) \nonumber\\
\times\delta(x_1-x_2+x_3-x_4)e^{-i\sum_{i<j}(-1)^{i+j+1}x_i\wedge x_j }.
\end{align}
These vertices are depicted on the figure \ref{fig:vertices}. Note
that additional overall factors must be taken into account. These are
indicated on the figure \ref{fig:vertices}.

\subsection{Computing the tadpole diagram.}

\begin{figure}[!htb]
  \centering
  \includegraphics[scale=1]{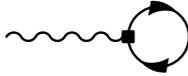}
  \caption[1-point]{\footnotesize{The non vanishing tadpole diagram.
      To simplify the figure, we do not explicitly draw all the
      diagrams that would be obtained from the vertices given on the
      figure 2 but indicate only the overall topology of the
      corresponding diagrams. Notice that the background lines are not
      explicitly depicted.}}
  \label{fig:1-point}
\end{figure}
It is instructive to present some relevant details of the calculation for the tadpole 1-point function
as an illustration of most of the technical features involved in the much more involved 
computation of the multipoint functions. The
amplitude for the tadpole depicted on figure \ref{fig:1-point} is \cite{PAPER1}
\begin{equation}
\mathcal{T}_1= \frac{1}{\pi^4\theta^4}\int d^4x\ d^4u\ d^4z\ A_\mu(u)\
e^{-i(u-x)\wedge z}\ C(x+z,x)\
((1-\Omega^2)(2\widetilde{x}_\mu+\widetilde{z}_\mu)-2\widetilde{u}_\mu).
\label{eq:tadpole1}
\end{equation}
This combined with the expression for the propagator
\eqref{eq:propag} yields
\begin{align}
\mathcal{T}_1=& \frac{\Omega^2}{4\pi^6\theta^6}\int d^4x\ d^4u\ d^4z
\int_0^\infty  \frac{dt\ e^{-tm^2}}{ \sinh^2(\widetilde{\Omega}t)
\cosh^2(\widetilde{\Omega}t)}\  A_\mu(u)\  e^{-i(u-x)\wedge z}\nonumber\\
&\times e^{-\frac{\widetilde{\Omega}}{4} (\coth(\widetilde{\Omega}t)z^2
+\tanh(\widetilde{\Omega}t)(2x+z)^2}
((1-\Omega^2)(2\widetilde{x}_\mu+ \widetilde{z}_\mu)-2\widetilde{u}_\mu).
\label{eq:tadpole2}
\end{align}
It is particularly usefull to reexpress \eqref{eq:tadpole2} under an expression that permits one to deal easily with the involved Gaussian integrals. For this purpose, it is convenient to introduce the following
8-dimensional vectors $X$, $J$ and the $8\times 8$ matrix $K$ defined
by
\begin{equation}
X=\begin{pmatrix} x\\ z \end{pmatrix}, \quad
K=\begin{pmatrix} 4\tanh(\widetilde{\Omega}t) \mathbb{I} &
2\tanh(\widetilde{\Omega}t)\mathbb{I} -2i\Theta^{-1} \\
2\tanh(\widetilde{\Omega}t)\mathbb{I} +2i\Theta^{-1} &
(\tanh(\widetilde{\Omega}t)+ \coth(\widetilde{\Omega}t))\mathbb{I}
\end{pmatrix} ,\quad
\ J=\begin{pmatrix} 0\\ i\tilde{u} \end{pmatrix}. \label{eq:tadpole2bis}
\end{equation}
The above trick \cite{PAPER1} can be adapted rather easily to simplify slightly the calculation of the higher order Green functions that will be presented in a while. Now, the combination of \eqref{eq:tadpole2bis}
with \eqref{eq:tadpole2} gives rise to
\begin{align}
\mathcal{T}_1=& \frac{\Omega^2}{4\pi^6\theta^6}
\int d^4x\ d^4u\ d^4z \int_0^\infty
\frac{dt\ e^{-tm^2}}{
  \sinh^2(\widetilde{\Omega}t)\cosh^2(\widetilde{\Omega}t)}\
A_\mu(u) \nonumber\\
&\times e^{-\frac{1}{2}X.K.X+J.X}
((1-\Omega^2)(2\widetilde{x}_\mu+ \widetilde{z}_\mu)-2\widetilde{u}_\mu).
\end{align}
By performing the Gaussian integrals on $X$, one then finds
\begin{equation}
\mathcal{T}_1=-\frac{\Omega^4}{\pi^2\theta^2(1+\Omega^2)^3}
\int d^4u \int_0^\infty  \frac{dt\ e^{-tm^2}}{
\sinh^2(\widetilde{\Omega}t)\cosh^2(\widetilde{\Omega}t)}\
A_\mu(u)\widetilde{u}_\mu\ e^{-\frac{2\Omega}{\theta(1+\Omega^2)}
\tanh(\widetilde{\Omega}t)u^2}. \label{eq:tadpole3}
\end{equation}
Finally, by inspecting the behaviour of \eqref{eq:tadpole3} for $t \to
0$, one realises that a quadratic as well as a
logarithmic UV divergence show up. Indeed, by performing a Taylor expansion of
\eqref{eq:tadpole3}, one obtains
\begin{align}
\mathcal{T}_1 =& -\frac{\Omega^2}{4\pi^2(1+\Omega^2)^3}
\left( \int d^4u\ \widetilde{u}_\mu A_\mu(u)\right)\
\frac{1}{\epsilon}\ -\frac{m^2\Omega^2}{4\pi^2(1+\Omega^2)^3}
\left( \int d^4u\ \widetilde{u}_\mu A_\mu(u)\right)\ \ln(\epsilon)
\nonumber\\
 & -\frac{\Omega^4}{\pi^2\theta^2(1+\Omega^2)^4}
\left( \int d^4u\ u^2\widetilde{u}_\mu A_\mu(u)\right)\
 \ln(\epsilon)\ + \dots,\label{eq:tasdepaul}
\end{align}
where $\epsilon \to 0$ is a cut-off and the ellipses denote finite
contributions. The fact that the tadpole is (a priori) non-vanishing
is a rather unusual feature for a Yang-Mills type theory. This will be
discussed more closely in a while.\par

\subsection{The multi-point contributions.}
\begin{figure}[!htb]
  \centering
  \includegraphics[scale=1]{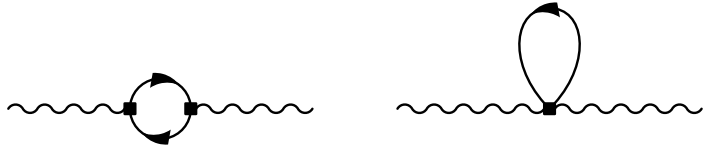}
  \caption[Two-point]{\footnotesize{Relevant one-loop diagrams
      contributing to the two-point function. To simplify the figure,
      we do not explicitly draw all the diagrams that would be
      obtained from the vertices given in figure 2 but indicate
      only the overall topology of the corresponding diagrams. Notice
      that the background lines are not explicitly depicted. The
      leftmost (resp.\ rightmost) diagram corresponds to the
      contribution $\mathcal{T}_2'$ (resp.\ $\mathcal{T}_2''$).}}
  \label{fig:2-point}
\end{figure}

The 2, 3 and 4-point functions can be computed \cite{PAPER1} in a somewhat similar 
way to the one given above but the calculations are much more cumbersome. The diagrams contributing to the 2-point function are depicted on the figure \ref{fig:2-point}. The respective contributions are found to be \par
\begin{subequations}
\begin{align}
\mathcal{T}_2' &= \frac{(1{-}\Omega^2)^2}{16\pi^2(1{+}\Omega^2)^3}
\left(\int \! d^4u\ A_\mu(u)A_\mu(u)\right) \frac{1}{\epsilon}
+ \frac{m^2(1{-}\Omega^2)^2}{16\pi^2(1{+}\Omega^2)^3}
\left(\int \! d^4u\ A_\mu(u)A_\mu(u)\right) \ln(\epsilon) \nonumber\\
 & +\frac{\Omega^2(1{-}\Omega^2)^2}{4\pi^2\theta^2(1{+}\Omega^2)^4}
\left(\int \!d^4u\ u^2A_\mu(u)A_\mu(u)\right) \ln(\epsilon)
\nonumber
\\
&- \frac{\Omega^4}{2\pi^2(1{+}\Omega^2)^4}\left(\int \!d^4u\
 (\widetilde{u}_\mu A_\mu(u))^2\right) \ln(\epsilon) \nonumber\\
& -\frac{(1{-}\Omega^2)^2(1{+}4\Omega^2{+}\Omega^4)}{
96\pi^2(1{+}\Omega^2)^4}
\left(\int \! d^4u\ A_\mu(u)\partial^2 A_\mu(u)\right) \ln(\epsilon)
\nonumber\\
& -\frac{(1{-}\Omega^2)^4}{96\pi^2(1{+}\Omega^2)^4}
\left(\int \! d^4u\ (\partial_\mu A_\mu(u))^2\right)\ln(\epsilon)+ \dots
\\
\mathcal{T}_2'' &= -\frac{1}{16\pi^2(1{+}\Omega^2)}
\left(\int \!d^4u\ A_\mu(u)A_\mu(u)\right) \frac{1}{\epsilon}
- \frac{m^2}{16\pi^2(1{+}\Omega^2)}
\left(\int\! d^4u\ A_\mu(u)A_\mu(u)\right) \ln(\epsilon) \nonumber\\
 & -\frac{\Omega^2}{4\pi^2\theta^2(1{+}\Omega^2)^2}
\left(\int\! d^4u\ u^2A_\mu(u)A_\mu(u)\right) \ln(\epsilon)
 \nonumber\\
  & + \frac{\Omega^2}{16\pi^2(1{+}\Omega^2)^2}
\left(\int d^4u\ A_\mu(u)\partial^2 A_\mu(u)\right) \ln(\epsilon)
  + \dots
\end{align}
\end{subequations}
One important comment is in order here. It is known that the regularisation of the
diverging amplitudes must performed in a way that preserves gauge
invariance of the most diverging terms for which, in $D$$=$$4$, quadratic UV divergences show up. One convenient regularisation \cite{PAPER1} is achieved by simply putting one cut-off $\epsilon$ 
on each integral over the Schwinger parameters, namely $\int_{\epsilon}^\infty dt$ while suitably tuning each cut-off to maintain gauge invariance. Here, this can be achieved with $\int_{\epsilon}^\infty
dt$ for $\mathcal{T}_2''$ while for $\mathcal{T}_2'$ the regularisation must be performed with $\int_{\epsilon/4}^\infty$. For more details, see \cite{PAPER1} or the talk of de Goursac at this conference.\par

\begin{figure}[!htb]
  \centering
  \includegraphics[scale=1]{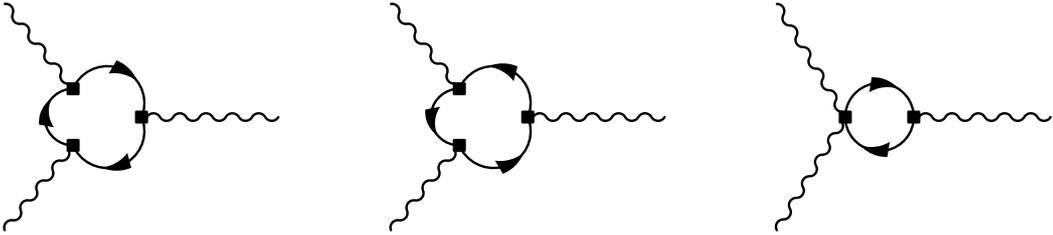}
  \caption[3-point]{\footnotesize{Relevant one-loop diagrams
      contributing to the 3-point function. Comments similar to those
      related to the figure 4 apply. The rightmost (resp.\ two
      leftmost) diagram(s) corresponds to the contribution
      $\mathcal{T}_3''$ (resp.\ $\mathcal{T}_3'$).}}
  \label{fig:3-point}
\end{figure}
The contributions corresponding to the diagrams of figure \ref{fig:3-point} are found \cite{PAPER1} to be 
\begin{subequations}
\begin{align}
\mathcal{T}_3' &=
\frac{\Omega^2(1{-}\Omega^2)^2}{8\pi^2(1{+}\Omega^2)^4}
\left( \int \! d^4u\ \widetilde{u}_\mu A_\nu(u)\{A_\mu,A_\nu\}_\star(u)
\right)\ln(\epsilon) \nonumber\\
 & +\frac{(1{-}\Omega^2)^2(1{+}4\Omega^2{+}\Omega^4)}{
48\pi^2(1{+}\Omega^2)^4}\left(\int \!d^4u\ (
(-i\partial_\mu A_\nu(u))[A_\mu,A_\nu]_\star(u)+\frac{4}{\theta^2} )\right)\ln(\epsilon)
+ \dots \\
\mathcal{T}_3'' &= -\frac{\Omega^2}{8\pi^2(1{+}\Omega^2)^2}
\left( \int \!d^4u\ (\widetilde{u}_\mu A_\nu(u)\{A_\mu,A_\nu\}_\star(u)
+\frac{4}{\theta^2})\right)\ln(\epsilon) \nonumber\\
 & +\frac{i\Omega^2}{8\pi^2(1{+}\Omega^2)^2}
\left(\int \!d^4u\ (\partial_\mu A_\nu(u))[A_\mu,A_\nu]_\star(u)
\right)\ln(\epsilon)+ \dots
\end{align}
\end{subequations}
\begin{figure}[!htb]
  \centering
  \includegraphics[scale=1]{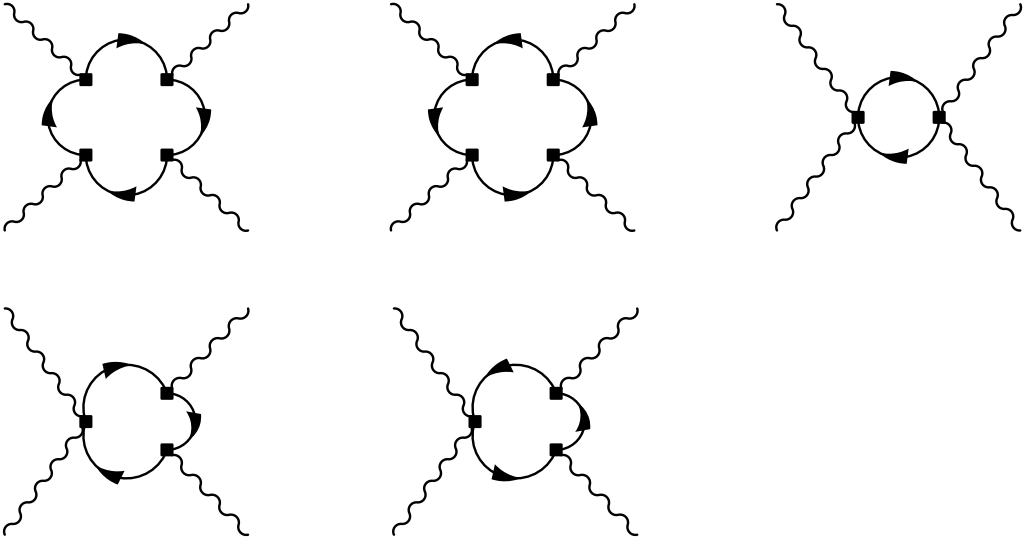}
  \caption[4-point]{\footnotesize{Relevant one-loop diagrams
      contributing to the 4-point function. Comments similar to those
      related to the figure 4 apply. Among the upper figures, the
      rightmost figure (resp.\ the two leftmost) diagram(s)
      corresponds to the contribution $\mathcal{T}_4'''$ (resp.\
      $\mathcal{T}_4'$). The lower diagrams correspond to
      $\mathcal{T}_4''$. }}
  \label{fig:4-point}
\end{figure}
In the same way, the 4-point contributions depicted on the figure
\ref{fig:4-point} are given by \cite{PAPER1}
\begin{subequations}
\begin{align}
\mathcal{T}_4' &= -\frac{(1{-}\Omega^2)^4}{96\pi^2(1{+}\Omega^2)^4}
\left( \int \!d^4u\ \left( (A_\mu\star A_\nu (u))^2
+2(A_\mu\star A_\nu(u))^2\right) \right) \ln(\epsilon)+ \dots
\\
\mathcal{T}_4'' &= \frac{(1{-}\Omega^2)^2}{16\pi^2(1{+}\Omega^2)^2}
\left(\int \!d^4u\ (A_\mu\star A_\mu(u))^2 \right) \ln(\epsilon)+ \dots
\\
\mathcal{T}_4''' &= -\frac{1}{32\pi^2}\left(\int \! d^4u\
(A_\mu\star A_\mu(u))^2 \right) \ln(\epsilon)+ \dots
\end{align}
\end{subequations}
\subsection{The effective gauge theory action}
Finally, by collecting the various contributions given above, we find
that the effective action $\Gamma(A)$ can be written as
\begin{align}
\Gamma(A) &= \frac{\Omega^2}{4\pi^2(1{+}\Omega^2)^3}
\left(\int \!d^4u\ (\mathcal{A}_\mu\star\mathcal{A}_\mu
-\frac{1}{4}\widetilde{u}^2) \right)
\left(\frac{1}{\epsilon}+m^2\ln(\epsilon)\right) \nonumber
\\
& -\frac{(1{-}\Omega^2)^4}{192\pi^2(1{+}\Omega^2)^4}
\left(\int \!d^4u\ F_{\mu\nu}\star F_{\mu\nu}\right)
 \ln(\epsilon) \nonumber\\
 & +\frac{\Omega^4}{8\pi^2(1{+}\Omega^2)^4}
\left(\int \! d^4u\ (F_{\mu\nu}\star  F_{\mu\nu}
+\{\mathcal{A}_\mu, \mathcal{A}_\nu\}_\star^2
-\frac{1}{4}(\widetilde{u}^2)^2)\right)\ln(\epsilon)+ \dots ,
 \label{eq:zegamma}
\end{align}
where $\mathcal{A}_\mu(u)= A_\mu(u)+{{1}\over{2}}\widetilde{u}_\mu$ and
$F_{\mu\nu}=\partial_\mu A_\nu-\partial_\nu
A_\mu-i[A_\mu,A_\nu]_\star$. To put the effective action into the form
\eqref{eq:zegamma}, it is convenient to use the following formulae
\begin{subequations}
\begin{align}
\int \!d^4x\ {\cal{A}}_\mu\star{\cal{A}}_\mu
&=\int \!d^4x(\frac{1}{4}{\widetilde{x}}^2+{\widetilde{x}}_\mu A_\mu
+A_\mu A_\mu),
\\
\int \!d^4x\ F_{\mu\nu}\star F_{\mu\nu}
&= \int d^4x\big(\frac{16}{\theta^2}-2(A_\mu\partial^2A_\mu+(\partial_\mu A_\mu)^2)
-4i\partial_\mu A_\nu[A_\mu,A_\nu]_\star-[A_\mu,A_\nu]_\star^2 \big),
\\
\int \!d^4x\; \{{\cal{A}}_\mu,{\cal{A}}_\nu\}^2_\star
&=\int \!d^4x\; \big(\frac{1}{4}({\widetilde{x}}^2)^2
+2{\widetilde{x}}^2{\widetilde{x}}_\mu A_\mu
+4({\widetilde{x}}_\mu A_\mu)^2+2{\widetilde{x}}^2A_\mu A_\mu
\nonumber
\\
&+2(\partial_\mu A_\mu)^2
+4{\widetilde{x}}_\mu A_\nu\{A_\mu,A_\nu\}_\star
+\{A_\mu,A_\nu\}^2_\star\big).
\end{align}
\end{subequations}

Let us now discuss the main features of this effective action together with some of the corresponding implications.

\section{Discussion.}

Field theories defined on Moyal spaces \cite{Douglas:2001ba,Wulkenhaar:2006si} appear so far to be interesting prototypes of field theories defined on "noncommutative spaces" whose specific classical (and hopefully quantum) features should be conveniently handled within a framework of noncommutative geometry \cite{CONNES,CM}. Within this latter algebraic framework, the commutative algebras of functions on differentiable manifolds are replaced by associative algebras that plays the role of "noncommutative spaces".  One example, among many others, is provided by the associative Moyal algebras that have been considered in the mathematical literature a long time ago from various viewpoints (see e.g. \cite{Grossmann:1968}, \cite{Gracia-Bondia:1987kw,Varilly:1988jk} and references therein). In noncommutative geometry, the relevant configuration spaces for field theories are modules over these associative algebras, interpreted as noncommutative analogues for the set of sections of vector bundles, while the linear spaces of the derivations of these algebras can be viewed as the noncommutative counterparts of the spaces of vector fields of commutative spaces. This permits one to define natural noncommutative extensions of connections and curvatures \cite{CONNES,CM}. The particular situation considered in this talk corresponds to the free module case, the most widely studied case in the literature, where the algebra ${\cal{M}}$ is viewed as a hermitian module over itself. Then, the
noncommutative analogue of gauge transformations are associated with the automorphisms of hermitian modules. The fact that ${\cal{M}}$ has only inner derivations implies the existence of a canonical gauge-invariant connection \cite{Dubois-Violette:1989vq, Masson:1999} from which a natural tensorial form can be defined. It is by the way related to the so-called covariant coordinates \cite{Douglas:2001ba}.\par
While several renormalisable noncommutative matter field theories have now been clearly identified, thanks to the "harmonic solution" proposed initially in \cite{Grosse:2004yu}, the construction of renormalisable actions for gauge theories defined on noncommutative Moyal spaces remains so far a challenging problem. One important step toward its solution is to guess possible form(s) for a candidate as a renormalisable gauge action and in particular a possible expression for the gauge counterpart (if it exists at all) of the harmonic term of \cite{Grosse:2004yu}. This is what we actually obtained in \cite{PAPER1}.\par
It is known from general arguments of quantum field theory that a valuable information about the possible form for gauge actions can be extracted from those effective gauge actions obtained by integrating over the matter degrees of freedom of a matter action coupled in a gauge invariant way to an external gauge potential. So, we started from, says $S(\phi,A)$, a renormalisable complex-valued scalar field theory with harmonic term \cite{Grosse:2004yu,Gurau:2005gd}, minimally coupled to an external gauge potential. We computed at the one-loop order the effective gauge theory action $\Gamma(A)$ \cite{PAPER1} given in \eqref{eq:zegamma}, obtained by integrating over the scalar field $\phi$, for any value of the harmonic parameter $\Omega \in [0,1]$ in $S(\phi,A)$. Independently of our analysis,  H. Grosse and M. Wohlgennant have recently carried out a nice calculation within the matrix-base \cite{Grosse:2007hh} of the one-loop effective gauge action obtained from a scalar theory with harmonic term different from ours, extending a previous work \cite{Grosse:2006hh} dealing with the limiting case $\Omega=1$. The scalar theory considered in \cite{Grosse:2007hh} is somehow similar to the one that one would obtain by considering "covariant derivative" as given in \eqref{eq:derivationcov} with symmetry transformations as those given by \eqref{eq:adjoint-de}. It can be easily verified that our effective action agrees globally with the one given in \cite{Grosse:2007hh}, up to unessential numerical factors. For more details, see M. Wohlgennant's talk at this conference.\par
We find that the effective gauge theory action involves, beyond the usual expected Yang-Mills contribution
\mbox{$\sim \int d^4x\ F_{\mu\nu}\star F_{\mu\nu}$}, additional terms
of quadratic and quartic order in ${\cal{A}}_\mu$
\eqref{eq:arondvrai}, $\sim \int d^4x\
{\cal{A}}_\mu\star{\cal{A}}_\mu$ and $\sim \int d^4x\
\{{\cal{A}}_\mu,{\cal{A}}_\nu\}_\star^2$. These additional terms are
gauge invariant thanks to the fact that, as a tensorial form, ${\cal{A}}_\mu$ transforms 
covariantly under gauge transformations. The quadratic term involves a mass term for the
gauge potential $A_\mu$. Recall that such a bare mass term for a gauge
potential would be forbidden by gauge invariance in commutative Yang-Mills
theories. At this point, one important comment relative to \eqref{eq:zegamma} is
in order. The fact that the tadpole is non-vanishing (see \eqref{eq:tasdepaul}) is a rather unusual feature for a Yang-Mills type theory. This non-vanishing implies automatically the occurrence
of the mass-type term $\int d^4x\ {\cal{A}}_\mu\star{\cal{A}}_\mu$ as
well as the quartic term $\int d^4x\ \{{\cal{A}}_\mu,{\cal{A}}_\nu\}_\star^2$.\par
It is tempting to conjecture that the following
class of actions ($\alpha$, $\Omega^\prime$, $\kappa$ and $g$ are parameters)
\begin{align}
S=\int d^4x \Big(\frac{\alpha}{4g^2}F_{\mu\nu}\star F_{\mu\nu}
+\frac{\Omega^\prime}{4g^2}\{{\cal{A}}_\mu,{\cal{A}}_\nu\}^2_\star
+\frac{\kappa}{2} {\cal{A}}_\mu\star{\cal{A}}_\mu\Big) \label{eq:decadix}
\end{align}
involves suitable candidates for renormalisable actions for gauge
theory defined on Moyal spaces. The naive action for a Yang-Mills theory on the Moyal space, $\sim \int d^4x\
F_{\mu\nu}\star F_{\mu\nu}$, taken alone, has UV/IR mixing
\cite{Hayakawa:1999yt,Matusis:2000jf}. In \eqref{eq:decadix}, the second term built from
the anticommutator may be viewed as a ``gauge counterpart'' of the
harmonic term. According to the above discussion, the presence of the
quadratic and quartic terms in ${\cal{A}}_\mu$ in \eqref{eq:decadix}
will be reflected in a non-vanishing vacuum expectation value for
$A_\mu$. The consequences of a possible occurrence of this non-trivial
vacuum remain to be understood in view of a subsequent BRST gauge-fixing of a gauge action derived from
\eqref{eq:decadix}. .\par

\vskip 1 true cm

\noindent
{\bf{Acknowledgements}}: 
I would like to thank the organiserss of the "International Conference on Noncommutative Geometry and Physics" for their kind invitation. Collaboration with A. de Goursac and R. Wulkenhaar is warmly acknowledged. I am also grateful to D. Blaschke, M. Dubois-Violette, H. Grosse.,  T. Masson and M. Wohlgennant for interesting discussions during the conference. This work has been supported by ANR grant NT05-3-43374 "GenoPhy". \par

\end{document}